\documentclass[aoas,MSNbibl,nameyear,rotating,seceqn,dvips]{arximspdf}
\usepackage{multirow}
\usepackage{graphicx}

\doi{10.1214/14-AOAS757} 
\volume{8}
\issue{3}
\pubyear{2014}
\firstpage{1800}
\lastpage{1824}
\docsubty{FLA}

\makeatletter
\def\mid{|}
\def\epsilon{\varepsilon}
\newcommand{\eqref}[1]{(\ref{#1})}
\def\E{\mathrm{E}}
\def\B{\mathcal B}
\def\R{\mathbb R}

\def\bX{\mathbf X}
\def\bZ{\mathbf Z}
\def\bY{\mathbf Y}
\def\bSigma{\bolds\Sigma}
\def\bone{\mathbf{ 1}}
\def\btheta{\bolds\theta}

\def\py{\mathrm y}
\def\px{\mathrm x}
\def\ga{\mathcal{GRF}}
\def\pp{\mathcal{PP}}
\def\gam{$\operatorname{Gamma}$}
\def\bmu{\bolds\mu}

\def\py{\mathrm y}
\def\px{\mathrm x}
\def\ga{\mathcal{GRF}}
\def\pp{\mathcal{PP}}
\def\gam{\operatorname{Gamma}}
\def\bmu{\bolds\mu}

\newproclaim{mydef}{Definition}
\newtheorem{mythm}{Theorem}

\makeatother

\begin{document}
\begin{frontmatter}

\title{A Bayesian hierarchical spatial point process model for
multi-type neuroimaging meta-analysis}
\runtitle{Bayesian multi-type meta-analysis}

\begin{aug}
\author[A]{\fnms{Jian}~\snm{Kang}\corref{}\ead[label=e1]{jkang30@emory.edu}\thanksref{m1}},
\author[B]{\fnms{Thomas E.}~\snm{Nichols}\ead[label=e3]{t.e.nichols@warwick.ac.uk}\thanksref{m2}},
\author[C]{\fnms{Tor~D.}~\snm{Wager}\ead[label=e4]{tor.wager@colorado.edu}\thanksref{m3}}\\
\and
\author[D]{\fnms{Timothy~D.}~\snm{Johnson}\ead[label=e2]{tdjtdj@umich.edu}\thanksref{T1,m4}}
\runauthor{Kang, Nichols, Wager and Johnson}
\thankstext{T1}{Supported by the NIH Grant 5-R01-NS-075066 (TDJ, TEN,
TDW), the United Kingdom's Medical Research
Council Grant G0900908 (TEN) and the Welcome Trust (TEN).
The work presented in this manuscript represents the views of the
authors and not necessarily that of the NIH,
UKMRC or the Welcome Trust.}
\affiliation{Emory University\thanksmark{m1},
University of Warwick\thanksmark{m2}, University of
Colorado\thanksmark{m3} and University of Michigan\thanksmark{m4}}
\address[A]{J. Kang\\
Department of Biostatistics and Bioinformatics\\
Department of Radiology and Imaging Sciences\\
Emory University\\
Atlanta, Georgia 30322\\
USA\\
\printead{e1}}
\address[B]{T.~E. Nichols\\
Department of Statistics \&\\
\quad Warwick Manufacturing Group\\
University of Warwick\\
Coventry CV47 7AL\\
United Kingdom\\
\printead{e3}}
\address[C]{T.~D. Wager\\
Department of Psychology and Neuroscience\\
University of Colorado\\
Boulder, Colorado 80309\\
USA\\
\printead{e4}}
\address[D]{T.~D. Johnson\\
Department of Biostatistics\hspace*{22pt}\\
University of Michigan\\
Ann Arbor, Michigan 48104\\
USA\\
\printead{e2}}
\end{aug}


\received{\smonth{8} \syear{2013}}
\revised{\smonth{5} \syear{2014}}

%
\begin{abstract}
Neuroimaging meta-analysis is an important tool for finding consistent
effects over studies that each usually have 20 or fewer subjects.
Interest in meta-analysis in brain mapping is also driven by a recent
focus on so-called ``reverse inference'': where as traditional
``forward inference'' identifies the regions of the brain involved in a
task, a reverse inference identifies the cognitive processes that a
task engages. Such reverse inferences, however, require a set of
meta-analysis, one for each possible cognitive domain. However,
existing methods for neuroimaging meta-analysis have significant
limitations. Commonly used methods for neuroimaging meta-analysis are
not model based, do not provide interpretable parameter estimates, and
only produce null hypothesis inferences; further, they are generally
designed for a single group of studies and cannot produce reverse
inferences. In this work we address these limitations by adopting a
nonparametric Bayesian approach for meta-analysis data from multiple
classes or types of studies. In particular, foci from each type of
study are modeled as a cluster process driven by a random intensity
function that is modeled as a kernel convolution of a gamma random
field. The type-specific gamma random fields are linked and modeled as
a realization of a common gamma random field, shared by all types, that
induces correlation between study types and mimics the behavior of a
univariate mixed effects model. We illustrate our model on simulation
studies and a meta-analysis of five emotions from 219 studies and check
model fit by a posterior predictive assessment. In addition, we
implement reverse inference by using the model to predict study type
from a newly presented study. We evaluate this predictive performance
via leave-one-out cross-validation that is efficiently implemented
using importance sampling techniques.
\end{abstract}

%
\begin{keyword}
\kwd{Bayesian spatial point processes}
\kwd{classification}
\kwd{hierarchical model}
\kwd{random intensity measure}
\kwd{neuorimage meta-analysis}
\end{keyword}
\end{frontmatter}

\section{Introduction}\label{sec:introduction}

Functional neuroimaging has experienced rapid growth since the early
nineties when Functional Magnetic Resonance Imaging (fMRI) was
developed. As the number of studies has grown exponentially, for
example, from
two fMRI studies in 1993 to over 2600 in 2012,\setcounter{footnote}{1}\footnote{Based on a
PubMed search for ``fMRI'' in the title
or abstract.} neuroscientists are increasingly interested in formal
synthesis of findings across studies via meta-analysis [\citet
{Yarkoni:2010}]. A neuroimaging meta-analysis mitigates the problems of
a single functional neuroimaging study. For example, most neuroimaging
studies have relatively low power due to small sample size. For
example, a recent meta-analysis of 90 neuroimaging studies on emotion
found that the median sample size was a mere 13 subjects [\citet
{Lindquist2012}]. Meta-analysis increases statistical power by combining
results from several smaller studies. Another problem is that many
published fMRI studies use ad hoc significance thresholds that result
in high false positive rates and idiosyncratic findings. Thus the
principal motivations for neuroimaging meta-analysis are to increase
statistical power and to find consistent activation regions across
studies, making it possible to separate reproducible findings from
idiosyncratic ones.

Another important motivation for meta-analysis is the recent interest
in ``reverse inference'' [\citet{Poldrack2011}]. A traditional fMRI
analysis conditions on the task paradigm and produces a ``forward
inference'', identifying the brain regions involved in the task. The
cognitive scientist will then display this set of brain regions and
argue, qualitatively, that this is evidence that the task produced a
particular cognitive state. However, the resulting brain regions may be
nonspecific and in fact activated by a range of cognitive tasks. In
one particularly egregious example [\citet{Iacoboni2007}], a
neuro-politics study inferred that brain activity in the anterior
cingulate, induced by images of Hillary Clinton, indicated that
subjects were experiencing emotional conflict; in fact, the anterior
cingulate is the most commonly active brain region, found in about 20\%
of all studies [\citet{Yarkoni:etal:2011}] that range from working memory
to decision making, as well as emotional processing.
Hence there is great interest in using predictive analyses to estimate,
conditional on brain activation map, the most likely class of task
paradigms that gave rise to the data. This process, referred to as
reverse inference, requires a set of meta-analyses, one for each class
of task paradigms. Reverse inference can also be used to test the
validity of the definition of task categories. That is, if studies can
be reliably classified between fine subdivisions of a task type, this
is evidence that the subdivisions are typified by unique patterns brain
activity and are not arbitrary constructs.

The information that is routinely reported in the literature, and thus
available for a meta-analysis, are the spatial locations of local
maxima of statistic values within each region of significant
activation. These locations are referred to as peak activation
locations, or foci. Functional neuroimaging meta-analysis studies based
on these foci are called coordinate based meta-analysis (CBMA). Many
CBMA methods have been proposed, including \citet
{Fox-1997,Nielsen-2002,Turkeltaub-2002,Wager-2004,Kober-2008,eick:lair:gref:2009,radua:2009,Kang:etal:2011}
and \citet{Yue:etal:2012}. To date, the most widely used methods are
kernel based
methods including activation likelihood estimation [\citet{Turkeltaub-2002},
ALE], modified ALE [\citet{eick:lair:gref:2009}, modALE]
and multilevel kernel density analysis [\citet{Kober-2008}, MKDA].
However, these methods have serious limitations. First, they require ad
hoc spatial kernel parameters, which must be pre-specified in the
analysis. Second, ALE maps are difficult to interpret, as they are
couched in probabilistic terminology but are not actually based on a
formal statistical model. Third, they are based on a massive univariate
approach that lacks an explicit spatial model. Thus, these methods can
only detect effects that consistently overlap in space, and cannot
assess spatial variability inherent in the foci.

To address these deficiencies \citet{Kang:etal:2011} proposed a Bayesian
hierarchical independent cluster process (BHICP) model.
This model is for a single class of studies, and does not accommodate
the multi-type point processes needed for reverse inference. While
BHICP model could be extended there are three significant
limitations: (1) the model involves many hyperprior distributions whose
parameters are challenging to elicit; (2) the posterior intensity
function is somewhat sensitive to the choice of hyperpriors; and (3) the
parametric form of the clustering may not be appropriate for all types
of spatial patterns found in CBMA data. Although it is possible to
extend this model by adding another level to the hierarchy, doing so
would only compound these problems.

More recently, \citet{Yue:etal:2012} proposed a Bayesian spatial
generalized linear model (SGLM) that treats the CBMA data as binary
random variables, one at each voxel. See \citet{Yue:etal:2012} for
details. There are several limitations to this approach as well. First,
this approach does not treat the individual studies as the units of
observation, but instead assumes the data at each voxel are the units
of observation. Second, the structure of CBMA data implies that the
number and locations of the foci within each study is random and this
approach does not respect this structure. Third, it is not a
generative, or predictive, model. While this SGLM approach does have
its merits, in this article we adopt the spatial point process
approach. The spatial point process approach more accurately captures
the stochastic structure of the data. Specifically that one unit of
data is an individual study comprised of a random number of foci
occurring at random locations.

Although many parametric spatial point process models have been
proposed for the analysis of multi-type point patterns [\citet
{Moller-book-2003}] any specific parametric intensity function is
difficult to justify. Therefore, we propose a nonparametric Bayesian
model to fit multi-type (emotion) meta-analyses by extending the
Poisson/gamma random field (PGRF) model developed by \citet
{Wolpert:Ickstadt:1998} to a hierarchical PGRF (HPGRF) model. The PGRF
model is a Cox process [\citet{cox:1955}] in which the intensity function
is modeled nonparametrically as the convolution of a spatial kernel and
a gamma random field. This model has found widespread use due to its
robustness in intensity function estimation and its computational
efficiency
[\citet{Ickstadt:Wolpert:1999,Best:etal:2000,Best:etal:2002,Stoyan:Penttinen:2000,Niemi:Fernandez:2010,Woodard:etal:2010}]. Our generalization from the PGRF model to the HPGRF
model is analogous to the extension of the mixture of Dirichlet process
priors model to the hierarchical mixture of Dirichlet process priors
model [\citet{Teh:etal:2006}]. In particular, we consider each type of
spatial point pattern as a realization of a PGRF model where the gamma
random field for each type is a realization from a population level
gamma random field (hence the hierarchy or ``random effects''). The
random intensity functions for the different types are related, thus
allowing not only aggregation of points within a type, but aggregation
of points across types.
The proposed HPGRF model has the following advantages over the BHICP
model: (1) It is a nonparametric model which provides more flexibility
in estimating the intensity function (which is also an advantage over
other spatial point process models such as the log-Gaussian Cox process
and Markov random field models [\citet{Moller-book-2003}]). (2) It requires
fewer hyperparameters and is less sensitive to the prior specification.
(3) It jointly estimates multi-type point patterns, borrowing strength
across the subtypes.

Our motivating data set comes from a functional neuroimaging
meta-analysis of emotions [\citet{Kober-2008}]. Kober et al.
collected data from 219 functional neuroimaging studies on five
emotions (sad, happy, anger fear and disgust). We will use reverse
inference to assess evidence for one perspective on emotional
regulation. The ``constructionist view'' [\citet{Lindquist2012}] suggests
that the basic categories of emotion (fear, disgust, etc.) arise from
complex combinations of elemental information-processing operations
across the brain. By this view, regions like the amygdala might be
involved in all of the basic emotions, but to different degrees with
other areas depending on the emotion type. Thus, the constructionist
theory suggests that a hierarchical model, or ``random effects'' type
of model, is appropriate. Thus, the BHICP and the PGRF models are not
applicable as they only model a single emotion type. In particular,
neither approach can borrow strength, nor model correlation, across the
different emotions as suggested by the constructionist view.

The remainder of this article is organized as follows. In Section~\ref
{sec:model}, we present our HPGRF model for multi-type point patterns.
We outline the model in Section~\ref{sec:hpgrf_single}. In
Section~\ref{sub:sec:hpgrf} we provide a theorem detailing the
expectation and
covariance of the associated counting process within a sub-type and the
covariance of the counting processes between subtypes for any two
regions of interest in the sampling window. In Section~\ref
{sub:sec:aug} we present a data augmentation scheme and in Section~\ref
{sub:sec:L'evy} we present an inverse L{\'e}vy measure representation
of the augmented model. We assess model performance via simulation
studies in Section~\ref{sec:simulation} and analyze the emotions
meta-analysis data set in Section~\ref{sec:examples}. We conclude with
a brief discussion in Section~\ref{sec:disc}.


\section{The model}\label{sec:model}

In this section, we start with a short overview of spatial point
processes, which are very useful tools in the analysis of spatial point
patterns [\citet{Moller-book-2003}], then introduce our HPGRF model for
multi-type spatial point patterns motivated by the meta-analysis of
functional neuroimaging data. In this article, all the point patterns
are defined on $\B\subset\mathbb R^3$ where $\B$ represents the
human brain.

\subsection{Spatial point processes}
For our purposes, a spatial point process $\bY$ is a random countable
subset on the brain, $\B$. For a spatial point process, there is an
associated counting process, $N_{\bY}(A)$, that counts the number of
points of $\bY$ in (well-behaved) subsets $A \subseteq\B$. One of the
most important spatial point processes is the Poisson point process. A
Poisson point process is characterized by an intensity function: a
nonnegative function that is integrable on all bounded subsets of $\B
$. Since the brain is a bounded subset of $\mathbb R^3$, for our
purposes, integrability on $\B$ is sufficient. We will use $\lambda
(y)$, $y \in\B$, to denote the intensity function. A spatial point
process is a Poisson point process if and only if (1) for all $A
\subseteq\B$, $N_{\bY}(A)$ follows a Poisson distribution with mean
$\Lambda(A) = \int_A \lambda(y)\,dy$, and (2) conditional on $N_{\bY}(A)
=n$, all points in $\bY$, that is, $y_1,\ldots, y_n$, are independent and
identically distributed with density $\lambda(y)/\Lambda(\B)$.

The intensity function in a Poisson point process is a known
deterministic function. This limits its use and flexibility in modeling data.
Thus, \citet{cox:1955} introduced the doubly stochastic Poisson
process; commonly known now as the Cox process. The Cox process
generalizes the Poisson point process by allow the intensity function
to be a random intensity function. Suppose now that $\lambda(y)$ is a
random, nonnegative function that is integrable on $\B$. If,
conditional on $\lambda(y)$, the point process $\bY$ is a Poisson point
process, then marginally, $\bY$ is said to be a Cox process driven by
$\lambda$.

Many Cox processes have been introduced in the literature with various
modeling assumptions on the random intensity function, $\lambda$. Most
of these assume that $\lambda$ is a parametrized function. For example,
the log-Gaussian Cox process [\citet{Moller:etal:1998}] assumes $\ln
[\lambda(y)] = U(y)$ where $U(y)$ is a Gaussian process parametrized by
a mean, a marginal variance and a correlation function (also
parametrized). There is a vast literature on spatial point processes.
We refer the interested reader to but a few: \citet{Illian:2008},
\citet
{Moller:Waagepetersen:2007} and \citet{vanLieshout:Baddeley:2002}.

\begin{figure}[b]

\includegraphics{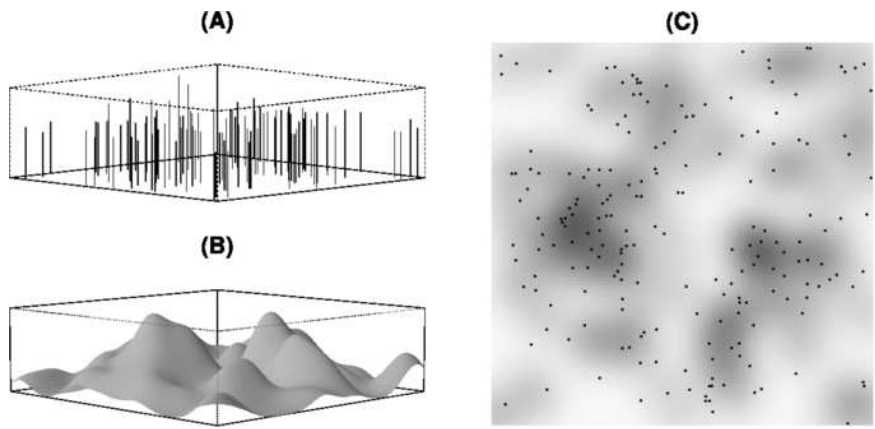}

\caption{A simulated two-dimensional gamma random field
[jump heights, panel \textup{(A)}], the corresponding intensity function
[convolution with a kernel, panel \textup{(B)}] and one
realization of the PGRF
[panel \textup{(C)}].}\label{fig:grf}
\end{figure}

As a nonparametric alternative to these parametric intensity
functions, \citet{Wolpert:Ickstadt:1998} proposed the Poisson/gamma
random field (PGRF) model. They model the random intensity function as
a convolution of a finite kernel, $k_{\sigma^2}(\py,\px)$, and a gamma
random field, $G(d\px)$: $\lambda(y) = \int_\B k_{\sigma^2}(\py
,\px)
G(d\px)$, where $\sigma^2$ is the kernel variance. As an example,
consider Figure~\ref{fig:grf}. In panel (A), we show the jump locations
and the jump heights from a simulated gamma random field on the unit
square. In panel (B), we show the intensity function produced by the
convolution of a Gaussian kernel with the gamma random field from panel
(A). Panel (C) shows the intensity function as an image with the points
representing a single realization of a point pattern drawn from
this PGRF. Note the distinctly non-Gaussian shapes in panels (B) and (C),
although the intensity function is modeled with a Gaussian kernel.

A gamma random field is characterized by a base measure $a(d\px)$ and
an inverse scale parameter $b$. If the random field $G(dx)$ is gamma
random field, we denote this by $G(dx) \sim\ga\{a(d\px), b\}$.
The gamma process (or random field) was first defined by \citet
{Ferguson:1973}. Formally, if a random field $G(d\px)\sim\ga\{a(d\px),
b\}$, then for any partition of the space $\B$, $\{A_k\}_{k=1}^K$ (i.e.,
$\B= \bigcup_{k=1}^K A_k$ and $A_k \cap A_j = \varnothing$ for
$k\neq
j$), $G(A_1),\ldots, G(A_K)$ are mutually independent and $G(A_k)$
follows a gamma distribution with shape $a(A_k)$ and inverse scale $b$.
\citet{Wolpert:Ickstadt:1998,Wolpert:Ickstadt:1998b} provide a
construction of a gamma random field that highlights the nonparametric
nature of the process (see, also, Section~\ref{sub:sec:L'evy}).

In this article, we assume that both $k_{\sigma^2}(\py, \px)$ and
$\lambda(\py)$ are Lebesgue measurable functions. We define
$K_{\sigma^2}(B, \px) = \int_{B} k_{\sigma^2}(\py, \px) \ell
(d\py)$ and
$\Lambda(B) = \int_{B} \lambda(\py) \ell(d\py)$ for any Lebesgue
measurable set $B \subseteq\B$, where $\ell$ is Lebesgue measure.
$K_{\sigma^2}( \cdot, \px)$ is called a kernel measure whereas
$\Lambda
(\cdot)$ is called an intensity measure. We can choose $k_{\sigma
^2}(\py, \px)$ to be a probability density function on $\B$. Thus,
we have
$\Lambda(d\py) = \int_\B K_{\sigma^2}(d\py,\px) G(d\px)$. The PGRF
model has been successful in the analysis of a single realization of a
point pattern, which is typical for most point pattern data. This model
enjoys most key advantages of parametric models, but can accommodate
irregular shapes of point clusters, is more flexible, and adaptive to data.

Before we introduce our model, one point of notation is in order. A
spatial Poisson point process is defined by specifying the sampling
window of interest (the brain, $\B$, in our case) and either an
intensity function or, equivalently, the associated intensity measure.
Both the intensity function and intensity measure carry the same
information about the process. We choose the latter to be consistent
with \citet{Wolpert:Ickstadt:1998}. Thus,
if $\bY$ is a Poisson point process on $\B$ with intensity measure
$\Lambda$, we denote this as $\bY\sim\pp\{\B, \Lambda(d\py)\}$ where
the differential $d\py$ is an infinitesimally small volume element in
$\B$.

\subsection{Hierarchical Poisson/Gamma random fields}\label{sec:hpgrf_single}

In this section, we generalize the PGRF model of \citet
{Wolpert:Ickstadt:1998} to model multi-type spatial point pattern with
between-type clustering or aggregation. For each emotion, the foci
reported from different studies are considered to be spatial point
patterns. Each spatial point pattern from each study, for a particular
emotion, is assumed to be an independent realization of a spatial point
process, where the spatial point processes underlying the different
emotions are dependent. We include this dependence between emotions as
it is suggested by the constructionist theory of emotion processing. We
model the dependence of different emotion-specific spatial point
processes using hierarchical gamma random fields.

Let $J$ denote the distinct emotion types studied and let $n_j$ denote
the number of independent studies of emotion $j$, $j=1,\ldots,J$.
Let ${\mathbf y}_{i,j}$, $i = 1,\ldots, n_j$, denote the
set of observed foci
from study $i$ of emotion $j$ and assume that each $
{\mathbf y}_{i,j}$ is a
realization from a Cox process, $\bY_{i,j}$, driven by a common random
intensity measure:
$
\Lambda_j(d\py)=\int_{\B} K_{\sigma^2_j}(d \py, \px) G_j(d \px)$,
where the $G_j(d\px)$ are independent and identically distributed with
common base measure $G_0(d\px)$ and inverse scale parameter~$\tau$. To
introduce dependence between types, we define $G_0(d \px)$ to be a
gamma random field with base measure $\alpha(d \px)$ and inverse scale
parameter $\beta$. In summary, our model is, for $ i = 1,\ldots,n_j,
j =
1,\ldots,J$,
%
\begin{eqnarray}\label{eq:HPGRF}
\bigl[\bY_{i,j}\mid\sigma_j^2,
G_j(dx) \bigr] &\stackrel{\mathrm { i.i.d.}} {\sim}& \pp \biggl\{\B,
\int_{\B} K_{\sigma^2_j}(d \py, \px) G_j(d
\px) \biggr\},
\nonumber
\\
\bigl[G_j(d \px) \mid G_0(d\px), \tau \bigr] &
\stackrel{\mathrm{ i.i.d.}} {\sim}& \ga\bigl\{G_0(d \px), \tau\bigr
\},
\\
\bigl[G_0(d \px) \mid\alpha(d\px), \beta \bigr] &\sim& \ga\bigl\{
\alpha(d \px), \beta\bigr\},
\nonumber
\end{eqnarray}
where the kernel variances, $\sigma_j^2$, base measure $\alpha(d\px)$
and inverse scale parameters $\tau$ and $\beta$, are all given
hyperprior distributions that we define later. Note that there are only
four parameters in this model---far fewer than the BHICP model. We note
here that the HPGRF generalizes the PGRF model of \citet
{Wolpert:Ickstadt:1998} in two ways. The first is trivial: we have
multiple observations of each emotion type. The second is trivial to
introduce, but is nontrivial algorithmically: we introduce another
level in the hierarchy. Thus, if we attempt to fit the PGRF model to
multi-type point patterns, then necessarily the multi-type patterns are
independent of one another. On the contrary, if we fit the HPGRF model
to multi-type point patterns, then the multi-type patterns are
dependent, as we now demonstrate.

\subsection{First and second moment properties}\label{sub:sec:hpgrf}
The HPGRF induces spatial correlation between the number of points in
any two regions of interest both within an emotion and between
emotions. This is an important aspect of our model for the emotion
meta-analysis data set and we will show in the simulation study section
that when there is aggregation of points between types that there is a
gain in efficiency as measured by the mean squared error.
We stress the point that by introducing this dependence between
intensity functions for the different emotions we take into account the
positive dependence (aggregation of points) between emotions offered by
the constructionist view of emotion processing. On the other hand, if a
repulsive process between emotion types is suggested, this model would
not be appropriate.

The conditional mean and covariance structure of $N_{\bY_{j}}(A)$ for
the HPGRF model are summarized in the following theorem whose proof is
given in Section~1 in the Web Supplementary Material~[\citet{Kang2014}].

%
\begin{mythm}\label{thm:property}
Within emotion type $j$ and for all $A, B\subseteq\B$,
%
\begin{eqnarray}
&&\E\bigl\{N_{\bY_{j}}(A) \mid\sigma^2_j,
\tau,\alpha,\beta\bigr\} =\frac
{1}{\tau\beta}\int_{\B}
K_{\sigma^2_j}(A, \px) \alpha(d\px ),\label
{eq:HPGRF_mean}
\nonumber
\\
\label{eq:HPGRF_cov1}
\qquad&&\operatorname{Cov}\bigl\{N_{\bY_{j}}(A), N_{\bY_{j}}(B)\mid
\sigma ^2_j,\tau,\alpha,\beta\bigr\}
\nonumber
\\[-8pt]
\\[-8pt]
\nonumber
&&\qquad= \frac{1}{\tau\beta}\int_{\B} K_{\sigma^2_j}(A\cap B,
\px) \alpha (d\px) + \frac{1+\beta}{\tau^2\beta^2}\int_{\B}
K_{\sigma
^2_j}(A, \px) K_{\sigma^2_j}(B, \px) \alpha(d\px).
\end{eqnarray}
Between emotion types $j$ and $k$ ($j\neq k$),
\begin{eqnarray}\label{eq:HPGRF_cov2}
&&\operatorname{Cov}\bigl\{N_{\bY_{j}}(A), N_{\bY_{k}}(B)\mid
\sigma^2_j, \sigma^2_k, \tau,
\alpha, \beta\bigr\}
\nonumber
\\[-8pt]
\\[-8pt]
\nonumber
&&\qquad= \frac{1}{\tau^2\beta^2}\int_{\B} K_{\sigma^2_j}(A,\px)
K_{\sigma
^2_k}(B,\px) \alpha(d\px).
\nonumber
\end{eqnarray}
\end{mythm}
This theorem shows that, as an a priori property, the intra-emotion and
inter-emotion number of points in $A$ and $B$ are positively
correlated, regardless of whether $A$ and $B$ are disjoint. When
$\sigma
^2_j = \sigma^2_k$, $j \neq k$, \eqref{eq:HPGRF_cov1} and \eqref
{eq:HPGRF_cov2} show that the intra-emotion covariance is larger than
the inter-emotion covariance.
Posterior inference of the HPGRF model is realized by the following
model representation.

\subsection{Data augmentation and complete data model}\label{sub:sec:aug}
Wolpert and
Ickstadt (\citeyear{Wolpert:Ickstadt:1998}) propose an alternative model
representation based on data augmentation that results in an efficient
MCMC algorithm for posterior estimation of the PGRF model. In this
section and the next, we generalize their approach to our hierarchical
model. First, we attach a mark to each point in $\bY_{j}$. Given our
model~\eqref{eq:HPGRF}, $N_{\bY_{j}}(\B)$ is a Poisson random variable
with mean $\Lambda_{j}(\B)$ and conditional on $N_{\bY_{j}}(\B)$,
$G_j(d\py)$ and $\sigma^2_j$, all points $Y_j \in\bY_j$ are independent
and identically distributed as
\begin{eqnarray*}
\bigl[Y_j \mid N_{\bY_{j}}(\B), G_j(d\py),
\sigma^2_j \bigr] &\stackrel {\mathrm{ i.i.d.}} {\sim}&
\Lambda_j(d \py)/ \Lambda_j(\B)\\
& = &\biggl(\int
_{\B} K_{\sigma^2_j}(d \py,\px) G_j(d \px)
\biggr)\Big/ \Lambda_j(\B).
\end{eqnarray*}
Next, for each $Y_j\in\bY_j$, we resolve this mixture distribution by
drawing an auxiliary random variable $X_j = \px_j\in\B$ from the
distribution,
\[
\bigl[X_j \mid Y_j = y_j,N_{\bY_{j}}(
\B), G_j(d \px),\sigma ^2_j \bigr] \sim
k_{\sigma^2_j}(y_j,\px) G_j(d \px) /
\lambda_j(y_j),
\]
where $\lambda_j(\py)$ is the intensity function of spatial point
process $\bY_j$. Lastly, define $(\bY_{j},\bX_{j}) = \{(Y_j, X_j), Y_j
\in\bY_j\}$. Then it is easy to show that $(\bY_j, \bX_j)$ is a
Poisson point process on $\B\times\B$ with intensity measure $
K_{\sigma
^2_j}(d \py,\px) G_j(d \px)$:
%
\begin{equation}
\bigl[(\bY_j,\bX_j) \mid K_{\sigma^2_j}(d \py,\px)
G_j(d \px ) \bigr] \sim \pp\bigl\{\B\times\B, K_{\sigma^2_j}(d
\py,\px) G_j(d \px)\bigr\} .\label
{eq:comp_model}
\end{equation}
By integrating out $\bX_j$, we recover the distribution of $\bY_j$ in
\eqref{eq:HPGRF}. It is the model in equation \eqref{eq:comp_model}
that we use in our posterior simulation which is based on the following
construction of a gamma random field.

\subsection{The L{\'e}vy measure construction}\label{sub:sec:L'evy}
Several methods have been proposed to simulate gamma random fields
including \citet{Bondesson:1982}, \citet{Damien:etal:1995} and \citet
{Wolpert:Ickstadt:1998b}. The inverse L{\'e}vy measure
algorithm [\citeauthor{Wolpert:Ickstadt:1998}
(\citeyear{Wolpert:Ickstadt:1998,Wolpert:Ickstadt:1998b})] provides an efficient
approach that has been successfully applied to the PGRF model. We
represent the algorithm in the following theorem.

%
\begin{mythm}\label{thm:L'evy}
Let $\theta_m \stackrel{\mathrm{ i.i.d.}}{\sim} \widetilde\alpha
(d\px)
\equiv\alpha(d\px)/\alpha(\B)$, $\nu_m = E_1^{-1}\{\zeta
_m/\alpha
^\prime(\theta_m)\}/\beta$, and $\zeta_m = \sum_{l=1}^m e_l$, for
$m =
1,2,\ldots,$ where $e_l \stackrel{\mathrm{ i.i.d.}}{\sim
}\operatorname{Exp}(1)$,
that is, the\break standard exponential distribution, and $E_1(t) = \int_t^{\infty} e^{-u}u^{-1}\,d u$. Let $\Gamma(d\px) =\break \sum_{m=1}^{\infty}
\nu_m \delta_{\theta_m}(d\px)$, then
\[
\Gamma(d\px) \sim\ga\bigl\{\alpha(d\px),\beta\bigr\}.
\]
If $\widetilde\alpha(d\px) = \widetilde\alpha^{\prime}(\px)\ell
(d\px)$,
then the joint distribution of $\{(\theta_m, \nu_m)\}_{m=1}^M$ has a
density with respect to $\prod_{m=1}^M \ell(d\theta_m)\ell(d \nu_m)$
proportional to
\[
\exp\bigl\{-E_1(\beta\nu_M)\widetilde
\alpha^{\prime}(\theta_M)\bigr\}\prod
_{m=1}^M\bigl[ \nu_m^{-1}\exp
\{-\nu_m \beta\}\widetilde\alpha^{\prime
}(\theta_m)
\bigr].
\]
\end{mythm}

\begin{figure}

\includegraphics{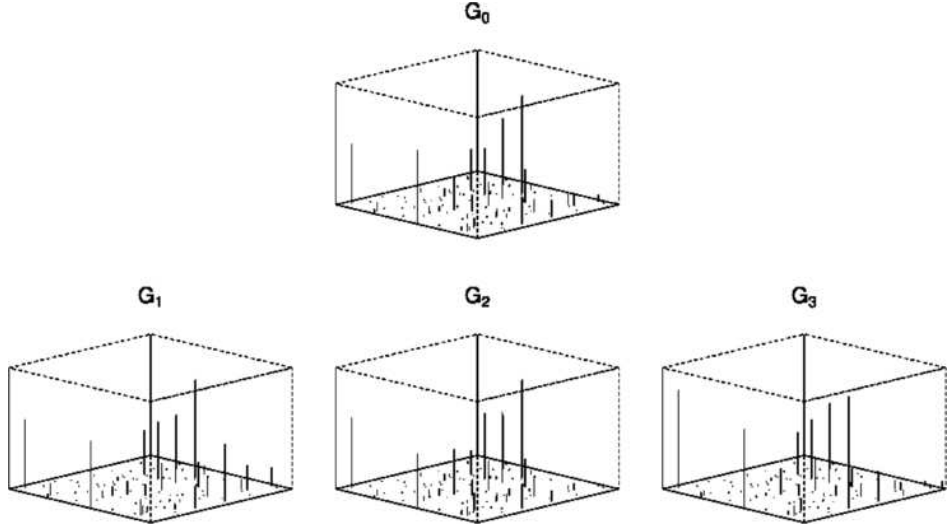}

\caption{Simulated two-dimensional hierarchical gamma
random fields, where $G_0$ is the population level gamma random field
and $G_j$ for $j = 1,2,3$ is the individual gamma random field. $G_0$
and all the $G_j$'s share the same support with different jump heights.
On average the jump heights of $G_j$ is about those of $G_0$.}\label{fig:hgrf}
\end{figure}

Note that $\widetilde\alpha(d\px)$ is a probability measure obtained
by normalizing $\alpha(d\px)$. The sequence $\{\zeta_m\}_{m=1}^M$
denotes the arrival times of the standard Poisson process on~$\R^+$.
The $\theta_m$ are the jump locations of the gamma random field while
$\nu_m$ is the jump height at location $\theta_m$. This is Theorem~1
and Corollary~2 of \citet{Wolpert:Ickstadt:1998} who also provide a
proof. Theorem~\ref{thm:L'evy} not only provides an efficient approach
to simulate from a gamma random field, it also provides an alternative
representation of a gamma random field that simplifies posterior
simulation. From this point forward, our L\'{e}vy measure construction
generalizes that of \citet{Wolpert:Ickstadt:1998} to Hierarchical
Poisson/Gamma random fields. Let InvL\'{e}vy$[\alpha(d\px), \beta]$ represent
the joint distribution of $\{(\theta_m, \nu_m)\}_{m=1}^\infty$ given
the base measure $\alpha(d\px)$ and inverse scale parameter $\beta$.
According to Theorem~\ref{thm:L'evy}, we can write:
%
\begin{equation}
\label{eq:Gamma_0} G_0(d\px) = \sum_{m=1}^{\infty}
\nu_m\delta_{\theta_m}(d\px),
\end{equation}
where $\{(\theta_m, \nu_m)\}_{m=1}^{\infty} \sim\mbox{InvL\'{e}vy}\{\alpha(d\px),
\beta\}$.
Note that $G_0$ has support on $\{\theta_m\}_{m=1}^{\infty}$. This
implies that each $G_j$ necessarily has the same support. See
Figure~\ref{fig:hgrf} for an illustration. Thus, there exist positive random
numbers $\mu_{j,m}, j = 1,\ldots, J$, such that
%
\begin{equation}
\label{eq:Gamma_j} G_j(d \px) = \sum_{m=1}^{\infty}
\mu_{j,m}\delta_{\theta_m}(d \px).
\end{equation}
Let $(B_1,\ldots, B_r)$ be any finite measurable partition of $\B$.
Let $A_l = \{m\dvtx\theta_m \in B_l\}$ for $l = 1,\ldots, r$. This implies
that $(A_1,\ldots, A_r)$ is a finite partition of the natural numbers.
For each $j$ and $l$, we have $G_j(B_l) \sim\gam(G_0(B_l), \tau)$ so
that $\sum_{m\in A_l} \mu_{j,m} \sim\gam (\sum_{m\in A_l} \nu_{m},
\tau )$. Thus, for $m = 1,2,\ldots,$
%
\begin{equation}
\label{eq:mu_jm} \mu_{j,m} \sim\gam(\nu_m, \tau).
\end{equation}
We note here that the $\mu_{j,m}$ are the jump heights of the gamma
random field $G_j(dx)$ and can be thought of as random effects about
the population level jump heights $\nu_m$ scaled by $\tau$. That is,
$E(\mu_{j,m}) = \nu_m/\tau$.
Finally, combining equations \eqref{eq:comp_model}, \eqref{eq:Gamma_0},
\eqref{eq:Gamma_j} and \eqref{eq:mu_jm} we have the following
equivalent representation of our HPGRF model:
%
\begin{eqnarray}\label
{eq:InvL'evy}
&&\bigl[(\bY_j,\bX_j) \mid\bigl\{(
\mu_{j,m},\theta_m)\bigr\} _{m=1}^{\infty},
\sigma^2_j \bigr]
\nonumber
\\
&&\qquad\sim \pp \Biggl\{\B \times\B, K_{\sigma^2_j}(d \py,\px) \sum
_{m=1}^{\infty} \mu _{j,m}\delta _{\theta_m}(d
\px) \Biggr\},
\nonumber
\\[-8pt]
\\[-8pt]
\nonumber
&& [\mu_{j,m} \mid\nu_m, \tau ] \stackrel{\mathrm{ i.i.d.}}
{\sim} \gam (\nu_m, \tau ),
\\
&&\bigl\{(\theta_m, \nu_m)\bigr\}_{m=1}^{\infty}
\sim\mbox{InvL\'{e}vy}\bigl\{\alpha(d\px), \beta\bigr\}.
\nonumber
\end{eqnarray}
In practice, we cannot sample $\{(\bY_j, \bX_j)\}_{j=1}^J$ according to
\eqref{eq:InvL'evy}, since it requires simulating an infinite number of
parameters which, in fact, reflects the nonparametric nature of both
the PGRF and the HPGRF models. Rather, we truncate the summation at
some large positive integer $M$. In the Web Supplementary
Material~[\citet{Kang2014}] we provide a theorem (Theorem~3) that states
we can approximate the conditional expectation of $N_{\bY_j}(A)$ to any
degree of accuracy we wish by a suitable choice of the truncation value
$M$. We also provide guidelines for choosing $M$ based on the inverse
scale parameters $\beta$ and $\tau$ and the base measure $\alpha
(\cdot
)$. After truncation, model \eqref{eq:InvL'evy} only involves a fixed
number of parameters which makes posterior computation straightforward.
We provide details of the posterior simulation algorithm in the Web
Supplementary Material [\citet{Kang2014}] as well.

\begin{table}[b]
\caption{Simulation study parameters for each of the
four aggregation regions}
\label{tab:pars} 
\begin{tabular*}{\textwidth}{@{\extracolsep{\fill}}lcccc@{}}
\hline
\textbf{Region} $\bolds{j}$ & \textbf{1} & \textbf{2} & \textbf{3} & \textbf{4}\\
\hline
$\sigma_j$ & 15 & 10 & 5 & 10\\
$\mu_j^\mathrm{T}$ & $(10,20)$ & $(70,30)$ &$(40,50)$ & $(60,75)$\\
$\Sigma_j$ & $\pmatrix{30& 15\cr 15& 15}$& $\pmatrix{30& -10\cr -10 &40}$
&$\pmatrix{20& -5\cr -5 &10}$ & $\pmatrix{10 &5\cr 5 &20}$ \\
$\epsilon$ & 0.001 & 0.001 & 0.001 & 0.001 \\
\hline
\end{tabular*}
\end{table}

\section{Simulation studies}\label{sec:simulation}

\begin{figure}

\includegraphics{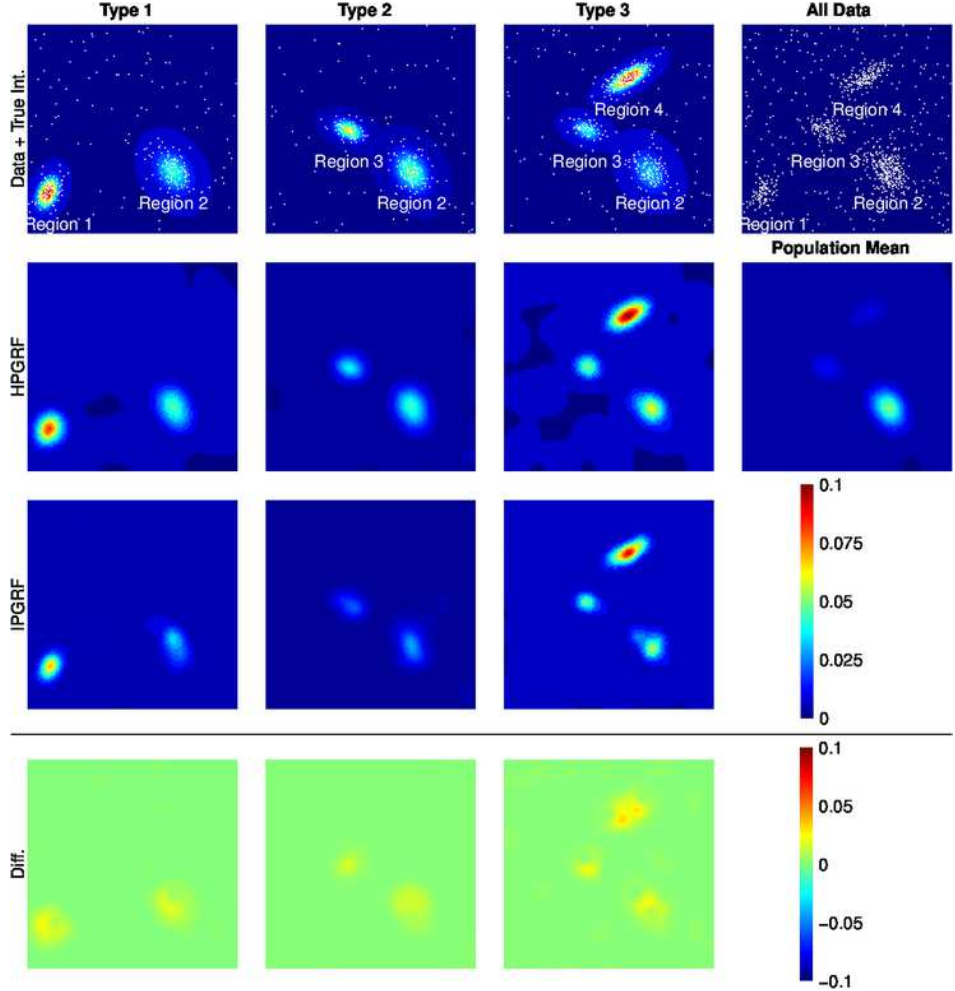}

\caption{Image intensities in the simulation
studies. Top row: True intensity functions with simulated data points
from one realization. 2nd row: Estimated posterior mean intensity
functions for the three types and for the population-level mean obtain
from our HPGRF model for one simulation. 3rd row: Estimated posterior
mean intensity functions obtained from the IPGRF model for one
simulation. Bottom row: Difference image (HPGRF--IPGRF).}\label
{fig:simulations}
\end{figure}

We simulate 2D spatial point patterns on a region $A = [0,100]^2$ from
three modified Thomas processes [\citet{vanLieshout:Baddeley:2002}]. Specifically, for $i
= 1,\ldots, N$ and $j = 1,2,3$, let $ [\bY_{i,j}\mid\bmu
,\bSigma
 ] \sim\pp\{A,\Lambda_j(d \px)\}$. For our simulation studies,
$\Lambda_j$ has associated intensity function $\lambda_j(\px) =
\epsilon+ \sum_{(\theta, \mu,\Sigma)\in(\btheta, \bmu,\bSigma
)_j}\theta\phi_2(\px;\mu, \Sigma)$ where $\phi_d(\px;\mu,\Sigma)$
denotes the $d$-dimensional Gaussian density at $\px$ with mean $\mu$
and covariance $\Sigma$; while $\epsilon$ is the homogeneous background
intensity and accounts for points that do not cluster or aggregate
(i.e., scatter noise and outliers). We set the intensity parameters (see
Table~\ref{tab:pars}) such that the point patterns from different types
aggregate on four regions.
The three sub-types have intensity functions (see Figure~\ref
{fig:simulations}):
\begin{eqnarray*}
\lambda_1(\px) &=& \epsilon+\theta_2
\phi_2(\px;\mu_2,\Sigma_2) +
\theta_3 \phi_2(\px;\mu_3,
\Sigma_3),
\\
\lambda_2(\px) &=& \epsilon+\theta_2 \phi_2(
\px;\mu_2,\Sigma _2)+ \theta _4
\phi_2(\px;\mu_4,\Sigma_4),
\\
\lambda_3(\px) &=& \epsilon+ \theta_1
\phi_2(\px; \mu_1,\Sigma _1) +
\theta_2 \phi_2(\px;\mu_2,
\Sigma_2) + \theta_3 \phi_2(\px;\mu
_3,\Sigma_3).
\end{eqnarray*}
All three types aggregate in region 2. Types 2 and 3 aggregate region
3. Only type 1 points aggregate in region 1 and only type 3 points
aggregate in region 4 (Figure~\ref{fig:simulations}). Figure A in the
Web Supplementary Material~[\citet{Kang2014}] shows marginal posterior
histograms of intensity functions evaluated at centers of regions 1--4.
This implies that the proposed method well assesses the posterior
variability of intensity functions.

For comparison we fit a spatial point process for each emotion by
extending the PGRF model to account for multiple independent
realizations. We refer to this model as the independent PGRF (IPGRF) model.
We simulate $K = 1000$ data sets according to the model specifications
in the previous section and fit each data set with the HPGRF model and
the IPGRF models, respectively. Figure~\ref{fig:simulations} shows the
true intensity functions (top row) along with the estimated posterior
mean intensity functions for one simulated data for both our HPGRF
model (middle row) and the IPGRF model (bottom row). From this figure
we see that both models do a good job, qualitatively, at reproducing
the true intensity function. However, the HPGRF intensity appears to
have regions of high intensity that are more elliptically shaped and
closer to the truth.

\begin{table}[b]
\caption{Simulation study results. Comparison of the
IMSE and the IWMSE summary measures for our HPGRF model and the IPGRF model}
\label{tab:acc}
%
\begin{tabular*}{\textwidth}{@{\extracolsep{\fill}}lcccc@{}}
\hline
& \multicolumn{2}{c}{\textbf{IMSE}} & \multicolumn{2}{c@{}}{\textbf{IWMSE}}\\[-6pt]
& \multicolumn{2}{c}{\hrulefill} & \multicolumn{2}{c@{}}{\hrulefill}\\
\textbf{Region} & \textbf{HPGRF (s.e.)} & \textbf{IPGRF (s.e.)} &
\textbf{HPGRF (s.e.)} & \textbf{IPGRF (s.e.)} \\
\hline
A & 175.94 (22.42) & 227.69 (25.57) & 11.15 (2.68) & 17.11 (2.60) \\
1 & \phantom{0}46.75 (10.24) & \phantom{0}50.88 (11.83) &
\phantom{0}3.94 (1.07) & \phantom{0}4.20 (1.04) \\
2 & 22.76 (6.76) & 52.59 (9.53) & \phantom{0}0.58 (0.19) & \phantom{0}1.72 (0.34) \\
3 & 19.70 (4.55) & 28.03 (5.86) & \phantom{0}0.75 (0.24) & \phantom{0}1.16 (0.26) \\
4 & \phantom{0}97.31 (16.95) & \phantom{0}90.47 (19.40) & 10.79 (2.38) & 10.02 (2.38) \\
\hline
\end{tabular*}
\end{table}

To quantify model performance, we compute the sub-type average
integrated mean square error (IMSE) and integrated weighted mean square
error (IWMSE) on region $A$. These quantities are defined,
respectively, as
\begin{eqnarray*}
\operatorname{IMSE} &=& \frac{1}{JK}\sum_{j=1}^J
\sum_{k=1}^K \int_{A}
\bigl[\widetilde \lambda_{jk}(\px) - \lambda_j(\px)
\bigr]^2 \,d \px,\label{eq:MSE}
\\
\operatorname{IWMSE} &=& \frac{1}{JK}\sum_{j=1}^J
\sum_{k=1}^K \int_{A}
\bigl[\widetilde \lambda_{jk}(\px) - \lambda_j(\px)
\bigr]^2\lambda_j(\px) \,d \px.\label
{eq:IWMSE}
\end{eqnarray*}
Here $\widetilde\lambda_{jk}(\px)$ is the type $j$ estimated posterior
mean intensity function in the $k$th simulation and $\lambda_j(\px)$ is
the true intensity function. The IWMSE gives more weight to regions
with a large true intensity. Table~\ref{tab:acc} summarizes the IMSE
and the IWMSE in different regions. Over the entire region $A$, the
IMSE and IWMSE are, respectively, 23\% and 35\% smaller under the HPGRF
model than under the IPGRF model. In region 2 (within the true 0.95
probability ellipse), where all three types aggregate, the IMSE and
IWMSE under the HPGRF model are 57\% and 66\% smaller than under the
IPGRF model. In regions 1 and 4, where no inter-type aggregation
occurs, both the HPGRF and IPGRF models give similar IMSE and IWMSE
results (Table~\ref{tab:acc}). Thus, when the different types of point
patterns aggregate on a common region, the HPGRF model provides more
accurate intensity estimates. When the types do not share any
clustering on a region, the HPGRF has comparable performance with the IPGRF.

\section{Application}\label{sec:examples}

\begin{figure}[b]

\includegraphics{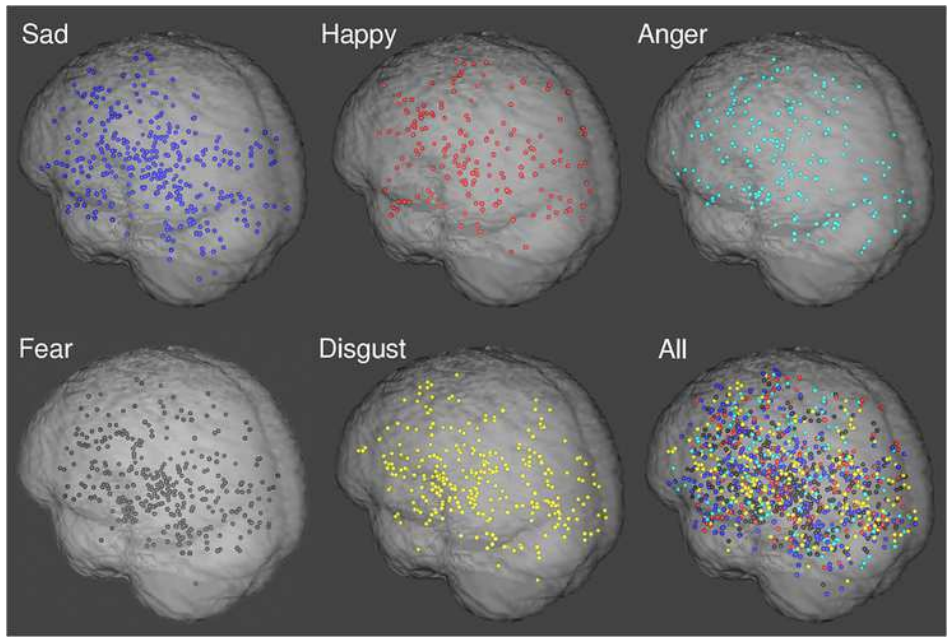}

\caption{Data: The 1393 foci reported from 219
studies of five emotions.}\label{fig:metadata}
\end{figure}

The emotion meta-analysis data set consists of 164 publications
designed to determine brain activation elicited by different emotions.
Researchers collected both fMRI and PET data. Many articles report
results from different statistical comparisons called ``contrasts'',
though we refer to each contrast as a ``study''. We use a subset of the
data, the 219 studies and 1393 foci for the five emotions sad, happy,
anger, fear and disgust. In Figure~\ref{fig:metadata} we display the
locations of all foci from the five emotions.

Recall that we have four priors to specify: $\alpha(dx)$, $\beta$,
$\tau
$, and $\sigma_j^2$. We assume $\alpha(dx)$, the base measure of
$G_0(dx)$, is Lebesgue measure. This implies that $\widetilde\alpha
(dx)$ in Theorem~\ref{thm:L'evy} has a uniform distribution over $\B$
and the jump locations $\theta_m$ of the gamma random fields are
uniformly distributed over $\B$, a priori. We assign the following
prior distributions to the hyper-parameters: $\sigma^{-2}_j\sim
U[0,10]$ and $\beta, \tau\sim\operatorname{Gamma}(2, 2)$. We
estimate the
posterior on 100,000 iterations of simulation after a burn-in of
20,000, saving every 50th iteration. We truncate the infinite summation
in the L\'{e}vy construction of the gamma random fields to $M =10\mbox{,}000$.
This results in a posterior estimated truncation error of 0.01. We
assess our model against both the IPGRF model and the BHICP model using
a posterior predictive check in Section~\ref{sec:PPA}.
We summarize results from a sensitivity analysis in Section~\ref
{sec:sens} with details in the Web Supplementary Material, Section~3.
We also summarize results from convergence diagnostics in Section~\ref
{sec:converge}.

We are interested in addressing the following questions. (1) Are there
consistent activation regions (aggregation of foci) across studies of
the same emotion? (2) Are there consistent activation regions across all
emotion types? (3) Can we accurately predict the emotion elicited in a
newly presented study?
For questions (1) and (2) we focus on the amygdalae which are bilateral,
almond-sized structures in the brain responsible for emotion processing
[\citet{Adolphs:1999}], especially anger and fear.

\begin{sidewaysfigure}

\includegraphics{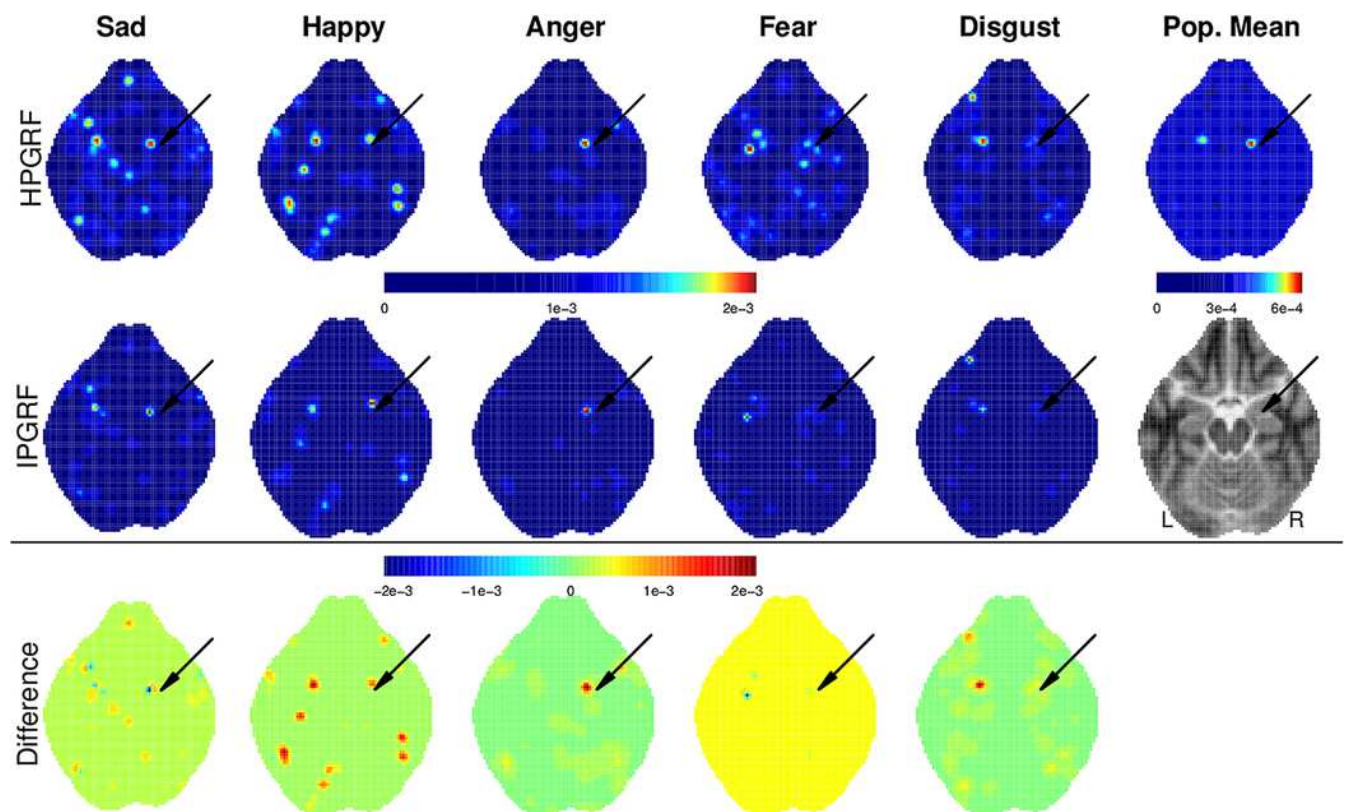}
\vspace*{-4pt}
\caption{Top row: A single axial slice ($Z =
-18$~mm) of the HPGRF posterior mean intensity estimates. (top row).
Middle row: The corresponding IPGRF estimates. The arrows point to the
right amygdala. All intensity functions have units of expected
foci/{mm}$^3$; the middle right image shows the corresponding
brain anatomy. Bottom row: Difference image (HPGRF--IPGRF).
Differences in this image can be clearly seen, especially the higher
intensity estimate from the HPGRF model in the right amygdala due to
borrowing of strength across the emotions.} \label{fig:meta_comp}
\end{sidewaysfigure}

(1) \textit{Are there consistent activation regions across studies of
the same emotion?}
For each emotion type we estimate the expected posterior intensity
function over the brain. We compare the intensity estimates between the
HPGRF and the IPGRF models for axial slice $\mathrm{Z}= -18$~mm (see
Figure~\ref{fig:meta_comp}). The intensity estimates are qualitatively
similar, however, the HPGRF intensity estimate appears more spatially
diffuse than the IPGRF intensity estimate. Furthermore, intensity
estimates from the HPGRF model tend to be larger than those from the
IPGRF model (Figure~\ref{fig:meta_comp}, 3rd row, where the difference between the HPGRF
and IPGRF intensities are shown). These observations are a direct
result of the fact that the jump locations of the gamma random fields
are shared across emotions, and hence there is a borrowing of strength
across the emotions.

All emotions show aggregation of foci, or consistent activation, in the
amygdalae, although to varying degrees. This basic finding is
consistent with previous meta-analytic summaries [\citet
{Costafreda2008,Lindquist2012}]. The posterior intensity is larger in
the left amygdala for fear and disgust, consistent with earlier
findings of overall left-lateralization in the amygdala [\citet
{Wager2003b}] and relative specificity for fear and disgust
[e.g., \citet{Costafreda2008,Wager2008,Lindquist2012,Yarkoni:2010,Yue:etal:2012}]. Sad and happy also show consistent activation in the
left amygdala as well, whereas sad, happy, anger and fear also show
strong right amygdala activation.

(2) \textit{Are there consistent activation regions across all emotion types?}
In line with the constructionist view of emotion processing, we are
interested in determining whether different emotions activate the same
areas of the brain, but perhaps, by varying degrees. This question can
be reformulated as a question of whether there is inter-type
(inter-emotion) aggregation of foci. To help answer this question we
define a ``population mean'' intensity measure. Recall that $\Lambda
_j(d\py)$ is the intensity measure for each point process $\bY_{i,j}$.
We define the\vspace*{1pt} ``population mean'' intensity measure by
$
\Lambda_0(d\py) =\tau^{-1}\int_{\B} \widetilde K(d\py, \px)
G_0(d\px)$,
where $\widetilde K(d\py,\px) = J^{-1}\sum_{j=1}^{J}K_{\sigma
^2_j}(d\py,\px)$. Thus, $\Lambda_0(d\py)$ is the average of the\vspace*{-1pt} expected
intensity measures of the different emotion types:
$\Lambda_0(d\py) = J^{-1} \sum_{j=1}^J E[\Lambda_j(d\py)\mid G_0,
\sigma
^2_j,\tau]$.

The image in the first row, last column of Figure~\ref{fig:meta_comp}
shows a slice of the posterior mean of the ``population level''
intensity. This slice intersects the amygdalae with the arrow pointing
to the right amygdala. There is relatively high intensity in the
amygdalae, confirming the importance of these brain structures in the
processing of emotions.

To measure the extent to which different emotions share common
activation regions we estimate the posterior correlations between the
different emotions in the amygdalae based on Theorem~\ref{thm:property}.
The average posterior correlations between the different emotions in
the left amygdala range from 0.69 to 0.71 and for the right amygdala
from 0.74 to 0.75. Thus, the data suggest that in each amygdala the
activation pattern among the five emotions are highly correlated. This
lends credence to the constructionist theory which attests that all
emotions elicit response in similar regions of the brain, but perhaps
to varying degrees, at least in the amygdalae.

(3) \textit{Can we accurately predict the emotion elicited in a newly
presented study?}
As described in the \hyperref[sec:introduction]{Introduction},
``reverse inference'' is used to
infer the most likely class of task to give rise to a particular study.
Such predictive inferences are straightforward with our HPGRF model.
Over the domain of five emotion subtypes, we can use a single study's
foci to make predictions about the emotion type of that study.

We compare our predictive method to previous work that combines the
MKDA and a na\"{i}ve Bayesian classifier (NBC) [\citet
{Yarkoni:etal:2011}]. For each study, this method creates binary
activation maps using the MKDA, with a value of 1 (activated) assigned
to each voxel in the brain if it is within a certain distance (a
spherical kernel size) of a reported focus, and 0 (nonactivated)
otherwise. These binary activation maps are in turn treated as feature
variables in the NBC. The study type, that is, the designed psychological
state, determines the class membership. Specifically, for each type, an
activation probability map is constructed by a weighted average of the
binary maps. Using the activation probability maps, the predictive
probability of the study type given activation from a new study is then
computed based on Bayes' theorem under an assumption of independent
voxels~[see \citet{Yarkoni:etal:2011} for details]. This method is very
computationally efficient and can handle extremely large sets of voxels
without difficulty. However, there are several potential drawbacks of
this method. First, NBC ignores the spatial dependence in the
activation maps, leading to biased predictive probabilities of the
class membership. Second, MKDA requires a fixed tuning parameter---the
kernel radius---that might affect classification performance: currently
MKDA simply fixes the kernel radius to some constant based on
experience rather than estimating it from data. Third, it only focuses
on the difference in the spatial distributions of foci between groups
while neglecting the absolute rates of foci, which may be important for
classification.

Our model has at least three advantages compared with the MKDA based
NBC. First, in the CBMA data, the number of foci and their locations
are random. Our HPGRF model explicitly models both the random number
and random locations of foci, as well as the spatial dependence between
foci. These features of the data are not modeled by the MKDA based NBC.
Second, our model is a more accurate representation of the true data
generating process, relative to how MKDA maps points to a voxel-wise
image with a spherical kernel. Third, our Bayesian model captures more
sources of variation and appropriately conveys the uncertainty in the
computation of the predictive probabilities that determines the classification.

We use Bayes' theorem to perform prediction using a leave-on-out cross
validation (LOOCV) approach. Details are given in the Web Supplementary
Material, Section~4.
We assume equal prior probability for each emotion, and fit our
Bayesian spatial point process classifier using the HPGRF model. As a
comparison, we also fit the classifier using the IPGRF model.
Table~\ref{tab:loocv} shows the LOOCV classifications rates based on our
HPGRF/IPGRF models as well as those based on the MKDA using the NBC.
Our spatial classifier correctly classifies 188 of the 219 studies, for
overall correct classification rate of $0.86 \pm0.024$ (mean $\pm$
standard error), far above random chance of 0.20. A simple average of
correct classification rates over emotions provides an average correct
classification rate of $0.85 \pm0.024$. The IPGRF based classifier
provides lower classification accuracy with the overall correct
classification rate of $0.75 \pm0.029$ and average correct
classification rate of $0.75 \pm0.029$. The MKDA based NBC (kernel
radius is 10~mm) correctly classifies 99 studies with an overall
correct classification rate of $0.45 \pm0.034$ and an average correct
classification rate of $0.36 \pm0.032$. Changing the MKDA kernel
radius to 5~mm, 15~mm and 20~mm did not improve the method's accuracy.

\begin{table}[t]
\caption{Confusion matrices of the LOOCV
classification. For the na\"{i}ve Bayesian classifier the overall
correct classification rate is $0.45 \pm0.034$ (mean $\pm$ s.e.) and
the average correct classification rate is $0.36 \pm0.032$. For the
independent Poisson/gamma random field model the overall correct rate
and average rate are both $0.75 \pm0.029$. For the hierarchical
Poisson/gamma random field model the overall correct classification
rate is $0.86 \pm0.024$ and the average rate is $0.85 \pm0.024$. The
largest standard error, based on the multinomial distribution, for any
of the methods for any emotion is 0.10}\label{tab:loocv}
\begin{tabular*}{\textwidth}{@{\extracolsep{\fill}}lcccccc@{}}
%
\hline
&  & \multicolumn{5}{c@{}}{\textbf{Confusion matrices}}\\[-6pt]
&  & \multicolumn{5}{c@{}}{\hrulefill}\\
&  \textbf{Truth}&\multicolumn{1}{c}{\textbf{Sad}} &
\multicolumn{1}{c}{\textbf{Happy}} &
\multicolumn
{1}{c}{\textbf{Anger}} & \multicolumn{1}{c}{\textbf{Fear}} &
\multicolumn{1}{c@{}}{\textbf{Disgust}}
\\
\hline
MKDA-NBC& sad & 0.38 & 0.11 & 0.07 & 0.40 & 0.04 \\
& happy & 0.11 & 0.25 & 0.03 & 0.56 & 0.06 \\
& anger & 0.12 & 0.23 & 0.00 & 0.50 & 0.15 \\
 & fear & 0.06 & 0.06 & 0.01 & 0.81 & 0.06 \\
& disgust & 0.09 & 0.16 & 0.05 & 0.32 & 0.39 \\[3pt]
IPGRF& sad & 0.78 & 0.09 & 0.04 & 0.07 & 0.02 \\
& happy & 0.00 & 0.81 & 0.03 & 0.17 & 0.00 \\
 & anger & 0.00 & 0.04 & 0.69 & 0.27 & 0.00 \\
& fear & 0.03 & 0.13 & 0.06 & 0.72 & 0.06 \\
& disgust & 0.02 & 0.09 & 0.02 & 0.09 & 0.77 \\[3pt]
%
HPGRF& sad & 0.91 & 0.04 & 0.00 & 0.04 & 0.00 \\
& happy & 0.00 & 0.83 & 0.08 & 0.08 & 0.00 \\
 & anger & 0.00 & 0.12 & 0.77 & 0.12 & 0.00 \\
& fear & 0.01 & 0.07 & 0.04 & 0.85 & 0.01 \\
&disgust & 0.02 & 0.05 & 0.02 & 0.02 & 0.89 \\
\hline
\end{tabular*}
\end{table}

Thus, our model based classifier does a good job at predicting the
emotion actually studied. The emotions anger and happy, when they are
misclassified, tend to be misclassified as fear. This finding
contradicts the simple assumption that similarity in our subjective
experience implies similarity in the brain processes that underlie
emotion. That assumption has driven psychologically based theories of
affect, such as the valence-arousal model \citet{Russell1999}, that
organize emotion based on direct experience. By contrast, our method
provides some early steps toward establishing taxonomies of emotion
based on similarity in brain activity patterns. Such taxonomies may be
based on properties that are identifiable at a psychological
level--for example, fear, anger, and happy all involve an aroused, activated
state, whereas disgust and sadness may not to the same degree--but they
need not respect our psychological distinctions. These taxonomies are
also relative to the level of analysis and spatial resolution one
considers: For example, neurons that encode negative and positive
information are intermixed in the amygdalae [\citet{Paton2006}], and thus
may elicit confusability between these types at the coarse
meta-analytic level of resolution. In short, the entire confusion
matrix provides information on the nature of emotion processing in the brain.

\subsection{Model assessment} \label{sec:PPA}

As a measure of model fit, we conduct a posterior predictive model
check using the $L$ function, a summary statistic for second order
properties of a point process [\citet{Baddeleyetal:2000,Illian:2009,Kang:etal:2011}]. The $L$ function can indicate aggregation or
clustering for a point process. For our model, $L(r; \cdot) =  \{3
K(r; \cdot)/4\pi \}^{1/3}$,
where
\[
K(r; \bY_{i,j}, \cdot) = \frac{1}{|\B|}\sum
_{y_1, y_2\in\bY_{i,j}} \frac{\bone[\|y_1 - y_2\| \leq r]}{\lambda_{1c}(y_1; \cdot)\lambda
_{1c}(y_2; \cdot)}.
\]
Consider the posterior predictive distribution of the differences
$\Delta_{i,j}(r) =  L(r; \bY_{i,j}, \cdot) - L(r; \bY^*_{i,j}, \cdot)$,
where\vspace*{1pt} $\bY^*_{i,j}$ is a simulated sample from the posterior predictive
distribution for study $i$ and type $j$. For a range of distances $r$,
if zero is an extreme value in the posterior predictive distribution of
$\Delta_{i,j}(r)$, then we question the fit of the model [\citet{Illian:2009,Kang:etal:2011}]. For $r>0$, we estimate the 95\% posterior
credible curves (as a function of $r$) of $\Delta_{i,j}(r)$, for each
the 219 studies. We consider zero an extreme value at a distance $r$ if
it lies outside of the 95\% posterior interval. We regard the model a
good fit for a study if zero is an extreme value in less than 10\% of
its range. For our HPGRF model, the model is a good fit for all 219
studies (100\%). For the IPGRF model, the model is a good fit for only
138 studies (63\%). Finally, for the parametric BHICP model, the model
is a good fit for only 142 studies (65\%). Thus, overall, our HPGRF
model provides a substantially better fit to the data based on this
posterior predictive assessment.

\subsection{Sensitivity analysis}\label{sec:sens}

We conduct a sensitivity analysis of the posterior intensity function
as a function of prior parameter distributions (for $\sigma_j^{-2}$,
$\beta$, $\tau$ and~$M$). We simulate the posterior using nine
different scenarios as shown in Table A in the Web Supplementary
Material~[\citet{Kang2014}], where Figure C presents one axial slice
($Z=-18$~mm) of the full 3D posterior mean intensity maps for different
scenarios. They looks qualitatively similar. Also, Table B in the Web
Supplementary Material~[\citet{Kang2014}] shows, for each emotion and for
the overall population, the minimum, median and maximum for the
expected posterior intensity function. These results show that the
posterior is not sensitive to the prior distributions over these nine scenarios.

\subsection{Convergence diagnostics and reproducibility}\label{sec:converge}

We also monitor convergence. We would like to monitor convergence of
the posterior intensity functions at each voxel. However, this is
impractical due to the extremely large number of voxels. Instead, we
run the model three separate times, with different random number
generation seeds and from over-dispersed starting values. From these
three runs we determine the location of the maximum difference in the
posterior intensity functions for each of the five emotions. We also
select ten other locations for which we monitor convergence. Some of
these locations are where the posterior intensity is larger and others
where it is small. These locations are chosen throughout the brain. We
also compute and save the integrated intensity functions (the posterior
expected number of foci for each study in each emotion type). We then
rerun the posterior simulation three more times, saving the posterior
draws of the intensity functions and using the Gelman--Rubin convergence
diagnostic for multiple chains [\citet{GelmanRubin:1992}]. We use a
burnin of 20,000 and run the chain for another 25,000 iterations and
save values at every 25th iteration for a total of 1000 saved draws
from the posterior (note that this is a smaller burnin period and a
short overall simulation than for the data analysis). The largest scale
reduction factor is 1.02 and the multivariate scale reduction factor is
1.09 [\citet{BrooksGelman:1998}]. A multivariate scale reduction factor
near 1 indicates convergence. From these results we are confident that
our chain has reached stationarity and that posterior estimates of
intensity functions can be reliably reproduced.

\section{Discussion}\label{sec:disc}

In this article, we propose a Bayesian nonparametric spatial point
process model, the HPGRF model, by generalizing the PGRF model
introduced by \citet{Wolpert:Ickstadt:1998} to hierarchical spatial
point process models. Our HPGRF model is appropriate for multi-type
spatial point pattern data when there is
aggregation between and within types. It accounts for positive
dependence in the point patterns both within and between types. That
is, it allows and models aggregation between points within types and
between points across types. Our model also allows for multiple,
independent, realizations of the spatial point process within each
type---as is demonstrated with the neuroimaging meta-analysis example
in Section~\ref{sec:examples}. We note here, that if there is repulsion
between types, such as when there is competition between species for
resources in ecological data, our model is not appropriate.

In our example analyses we provide ``population mean'' intensity
estimates to identify common regions that share clustering, or
aggregation, providing better interpretation of data. Results from the
emotion meta-analysis also lend support to the constructionist view of
emotions. The LOOCV results demonstrate that, at least for prediction
purposes, our model is more appropriate than the IPGRF model and
greatly outperforms a simple na\"{i}ve Bayes classifier. This
performance difference is evidence that the point process approach
captures important spatial and stochastic features of CBMA data. Such
classification results are also a first step in allowing users of fMRI
to make ``reverse inferences'' [\citet{Poldrack2011}].

The simulation studies shows that the HPGRF model improves intensity
estimate accuracy over the IPGRF model when aggregation is present
across types and does not suffer a loss of accuracy when the point
patterns arising from the different types are independent of one
another. Posterior predictive checks also indicate that our model fits
the data better than both the IPGRF and the BHICP models.
Sensitivity analyses and convergence diagnostics demonstrate that our
model is robust to prior specification and that posterior estimation of
the intensity function is reproducible.

Like the PGRF model, the HPGRF model can accommodate non stationary
processes by include spatially varying covariates. For example with a
single spatially varying covariate, $\bZ(\py)$ say, using a
semi-parametric regression approach, the random intensity measure for
each type $j$ can be modeled as $\tilde\Lambda_{j}(d\py; \bZ) =
\Lambda
_{j}(d\py) \exp \{\bZ(\py){\bolds\beta_j} \}$. We can
also assign an hierarchical prior on $\bolds\beta_j$ such that
the posterior estimates of $\bolds\beta_j$ borrow information
across different types. $\Lambda_j(d\py)$ is interpreted as the
baseline intensity measure and $\bolds\beta_j$ represents the
spatial covariate effect for type $j$.

There are several directions one can take to extend our model further.
First, the HPGRF model can be extended to more than two levels of
hierarchy to deal with more complex spatial point patterns. The depth
of the hierarchy would depend on the needs of the data analysis. For
instance, in the functional neuroimaging meta-analyses of emotion,
there are positive emotions (e.g., happy and affective) and negative
emotions (e.g., fear and disgust). One problem of interest is to
identify the common consistent activation regions for positive
emotions, negative emotions, and all emotions. This motivates the needs
for a third level in hierarchy: the first level models each type of
emotion; the second level models positive/negative emotions and the
third level models the entire population of emotions. Another
interesting extension is to allow the HPGRF model to accommodate
multiple dependent realizations of multi-type spatial point processes.
A practical use for such a model is the analysis of spatio-temporal
point pattern data, even for a single type.

Computationally, the L\'{e}vy construction relies on the truncation of
an infinite sum. The number of points, $M$, in the gamma random field
typically must be very large to achieve a reasonable level of accuracy
for the intensity estimates, thus the computation cost can be high. The
analysis present in Section~\ref{sec:examples} takes approximately 20
hours on a MAC Pro with 8~Gb of memory and a 3.0~GHz Intel processor.

There is potential to speed up the computation. One possible solution
is to approximate the gamma random field using a marked point process
according to the L\'{e}vy construction, where a point represents the
location of a jump and the mark is the height of the jump. Then we can
utilize the spatial birth--death process to simulate a gamma random
field with a random number of jumps. Currently, we are evaluating the
possibility of leveraging the relationship between the gamma process
and the Dirichlet process [\citet{Ferguson:1973}] and modifying one of
the many algorithms developed for Dirichlet process models [see, \citet
{Walker:2007}, e.g., which appears quite promising] for our HPGRF model.

Code is available by contacting the first author.

\section*{Acknowledgements}

We are grateful to Dr. Lisa Feldman Barrett, University Distinguished
Professor of Psychology, Northeastern University for providing the data
along with Tor Wager. We also thank the two reviewers, the Associate
Editor and the Editor for their insightful comments and constructive
suggestions that improve our manuscript substantially.

%
%


\begin{supplement}[id=suppA]
\stitle{Supplement to ``A Bayesian hierarchical spatial point process
model for multi-type neuroimaging meta-analysis''}
\slink[doi]{10.1214/14-AOAS757SUPP} 
\sdatatype{.pdf}
\sfilename{aoas757\_supp.pdf}
\sdescription{In this online supplemental article, we provide (1)
proofs of main theorems for the HPGRF model, (2) details on posterior
computations, (3) additional figures to assess the posterior
variabilities of intensity functions in simulation studies and data
application, (4) sensitivity analysis, and (5) details of a Bayesian
spatial point process classifier.}
\end{supplement}

\printaddresses
\end{document}